\documentclass[prd,nofootinbib,11pt]{revtex4}
\usepackage{amssymb,amsmath,latexsym,braket,fancyref,hyphenat}

\newcommand{\pd}{\partial}
\newcommand{\p}{\rho}
\newcommand{\hf}{\frac{1}{2}}
\newcommand{\pfv}{\rho_{\text{fv}}}
\newcommand{\ptv}{\rho_{\text{tv}}}
\newcommand{\psh}{\rho_{\text{max}}}
\newcommand{\Vo}{V(\rho_0)}
\usepackage{graphicx}
\usepackage{dcolumn}
\usepackage{bm}
\usepackage{placeins}

\numberwithin{equation}{section}
\usepackage{parskip}
\begin{document}

\begin{flushright}
MAN/HEP/2019/008\\
November 2019
\end{flushright}

\title{{\Large Goldstone Boson Effects on Vacuum Decay}\\[3mm] }
\begin{center}
 \begin{footnotesize}
 * Numerical results revised from the published version, conclusions remain unchanged.
\end{footnotesize}
\end{center}

\author {\large Mulham Hijazi$\,$\footnote{E-mail address: {\tt mulham.hijazi@postgrad.manchester.ac.uk}}}
\author {\large Apostolos Pilaftsis$\,$\footnote{E-mail address: {\tt apostolos.pilaftsis@manchester.ac.uk}}\\}

\affiliation{\vspace{0.2cm} 
  Consortium for Fundamental Physics, School
  of Physics and Astronomy, University of Manchester, Manchester, M13
  9PL, United Kingdom}

\begin{abstract}
\vspace{2mm}
\centerline{\small {\bf ABSTRACT} }
\vspace{2mm}\noindent

\noindent We study the effects of Goldstone modes on the stability of the vacuum in a $U(1)$ theory for a complex scalar field. The dynamics of the field resemble those of Keplerian motion in the presence of time-dependent friction, whose equations of motion imply a conserved quantity, $L$, reminiscent of conserved angular momentum. They also imply a persistent infinite barrier at $\rho=0$ and a divergent field value  at the origin of coordinates in flat spacetime, rendering any solution physically unattainable. However, in a spacetime punctured at the origin of coordinates, we find finite-action solutions to the equations of motion, which correspond to the size of the hole $a_0$, which in turn determines the tunneling point $\p_0$ and $L$. We find that the rates of vacuum decay get drastically enhanced by many orders of magnitude for all possible orderings in which the false and true vacua are placed in the potential. Finally, we show how Goldstone modes provide the necessary energy to overcome drag forces yielding finite-action solutions for any potential, including those that no such solutions exist for real scalar fields.

\end{abstract}

\maketitle

\vspace{10mm}
\section{Introduction}

Instantons in $D$-dimensions are $\text{O}(D)$ symmetric classical solutions of the equations of motion corresponding to a quantum tunnelling process of a scalar field through a potential that has more than one vacuum state. Classically, the transition from the false vacuum to the true vacuum of the theory is forbidden by energy conservation. However, the laws of Quantum Mechanics allow for such a process to occur through means of tunnelling. Energetics favour the tunnelling from the false vacuum to the true vacuum of the theory as a decay process, and the probability of such a decay depends on the value of the Euclidean action \cite{Coleman:1977py,Coleman:1978ae,Vainshtein:1981wh,Rajaraman:1982is,Shifman:1994ee,Linde:1981zj,Rubakov:1984bh,Rubakov:2002fi,Weinberg:2012pjx} .

In Quantum Mechanics, while adopting the well-known WKB approximation, consistent with the path integral formulation of quantum field theory, we find that the probability of tunnelling through a finite potential is proportional to $e^{iS}$, where $S$ is the action defined as the integral of the Lagrangian corresponding to a given potential  \cite{Ryder,Srednicki:2007qs,Peskin:1995ev}. By Wick rotating the time coordinate into Euclidean time $\tau\equiv-it$, and defining the Euclidean action as $S_E\equiv -iS$, we write the expression for the tunnelling rate as  \cite{Coleman:1977py}

\begin{align}
  \label{eq:Gamma}
\Gamma = A e^{-B}\ ,
\end{align}
where $A$ is a calculable prefactor, and $B=S_{E}(\phi_{\text{b}})- S_E(\phi_{\text{fv}})$, where $\phi_{\text{b}}$ and $ \phi_{\text{fv}}$ are the bounce and the false vacuum solutions, respectively. This means that only solutions that give a finite Euclidean action will lead to non-zero tunnelling rates.

The field-theoretic formulation of vacuum decay for a real scalar field in flat spacetime was first presented by Coleman and Callan in two separate papers~\cite{Coleman:1977py,Callan:1977pt}. The first of which introduced the thin wall approximation corresponding to potentials where the difference in potential energy between the false vacuum and the true vacuum is small compared to the height of the barrier, while the second aimed to calculate the prefactor $A$ by considering quantum corrections. These papers were followed by another paper by Coleman and De Luccia who studied gravitational effects on such decay~\cite{Coleman:1980aw}. In our paper, we will recap these formalisms, and extend them on complex scalar fields with non-zero Goldstone boson modes.

Goldstone bosons appear in theories which exhibit spontaneous symmetry breaking of continuous symmetries. They are necessarily massless and play an important role in physical phenomena such as endowing particles of the Standard Model (SM) with mass through the Higgs mechanism \cite{Goldstone:1962es}. Potentials which are independent of Goldstone fields yield the same vacuum expectation value (VEV) regardless of the evolution of the Goldstone field. However, Goldstone modes contribute to the rotational kinetic energy in the action. Therefore, our interest in this paper is to study the effects of non-zero Goldstone modes on the decay of the vacuum, particularly on the Euclidean action that appears in the exponent of (I.1), which determines the tunnelling rate.

The layout of this paper is as follows: Section~\ref{sec:Theory} presents theoretical preliminaries, starting with a subsection considering the flat spacetime case where we will highlight the impossibility of finding finite-energy solutions due to an unavoidable divergence at the origin of coordinates. The next subsection explores solutions in punctured spacetime which depend on the parameter determining the size of the hole and the value to which the field arrives after tunnelling. Such topological holes resemble primordial Einstein-Rosen wormholes which are solutions of the Einsteins field equations~\cite{Alonso:2017avz,Deng:2016vzb,Frampton:1976pb,Antoniou:2019awm}.

Section~\ref{sec:Numerics} explores numerical solutions for different potentials. We consider Coleman potentials where the energy difference between the false and true vacua is small compared to the height of the potential. We highlight the fact that the order in which the false and true vacua are placed within the potential makes a significant difference, as we find qualitatively distinct profiles of solutions for different orderings. Most importantly, we find that the tunnelling rates get enhanced by many orders of magnitude. We then consider Fubini potentials with a mass term, for which no tunnelling solution exists in flat spacetime. However, we find that Goldstone modes provide the necessary energy to overcome friction, and thus provide solutions to such problem. In the last Section~\ref{sec:Conclusions}, we summarise the results of our paper and briefly point out possible implications of such solutions in cosmology.

\section{Theoretical Background}\label{sec:Theory}

In this section we derive the equations of motion for a tunnelling field corresponding to a general action, first for the flat spacetime case, and then for the case of punctured curved spacetime.

\subsection{Euclidean flat spacetime}

We start our discussion by performing a Wick rotation to the action $S$, 
\begin{align}
S&=\int dt d^{D-1}x \bigg[\frac{1}{2}\pd_\mu\Phi^\dagger\pd^\mu\Phi-V(\Phi^\dagger\Phi)\bigg]\nonumber\\
&=i\int d\tau d^{D-1}x \bigg[\frac{1}{2}\pd_\mu\Phi^\dagger\pd_\mu\Phi+V(\Phi^\dagger\Phi)\bigg]\equiv i\int d\tau d^{D-1}x L_E \equiv iS_E \ ,
\end{align}
where $\tau\equiv -it$. By defining the evolution parameter $r\equiv \sqrt{\tau^2+|\vec{x}|^2}$ and using the relation for the volume element~\cite{Weisstein},
\begin{align}
dV_D=\frac{2 \pi^{\frac{D}{2}} r^{D-1}}{\Gamma(1+\frac{D}{2})}dr\ ,
\end{align}
we find that the Euclidean action can be written as
\begin{align}
S_E&= \frac{2 \pi^{\frac{D}{2}}}{\Gamma(1+\frac{D}{2})} \int_0^\infty dr \  r^{D-1}L_E=\frac{2 \pi^{\frac{D}{2}}}{\Gamma(1+\frac{D}{2})} \int_0^\infty dr \  r^{D-1}\bigg[\frac{1}{2}\pd_r\Phi\pd_r\Phi^\dagger+V(\Phi^\dagger\Phi)\bigg]\ .
\end{align}

To properly describe the tunnelling process, we assume that the field was trapped in the false vacuum at time $\tau_i=-\infty$, tunnels through the barrier to the turning point at time  $\tau=0$. However, by time reversal symmetry, we find that $\tau$ and $-\tau$ correspond to the same evolution parameter $r$. Therefore, as $r \rightarrow \infty$,  the field ``bounces'' 
back to the false vacuum state. The mechanism for tunnelling can be visualised as the materialisation, or the nucleation, of a bubble near the true vacuum at $r=0$, which grows at the speed of light. Far away from the origin, the vacuum is unperturbed \cite{Coleman:1977py}.

We write the complex scalar field as $\Phi=\rho e^{i\chi}$, where $\rho,\chi$ are real fields. 
The field $\chi$ is the Goldstone field. In terms of $\p$ and $\chi$, the Euclidean action becomes
\begin{align}
S_E= \frac{2 \pi^{\frac{D}{2}}}{\Gamma(1+\frac{D}{2})} \int_0^\infty dr \  r^{D-1}\bigg[\frac{1}{2}(\p')^2+\frac{1}{2}\rho^2(\chi')^2+V(\rho)\bigg]\ ,
\end{align}
where the primes denote differentiation with respect to $r$. From~\eqref{eq:Gamma}, it is clear that we must have a finite value for the Euclidean action in order for the tunnelling process to take place. For this to happen, we must have a vanishing potential as $r \rightarrow \infty$. Moreover, to conserve energy, the field must tunnel through the barrier with zero kinetic energy. Thus, we impose the condition $d\p/d\tau\rvert_{\tau=0}=0$ which translates to $d\p/dr\rvert_{r=0}=0$, since
\begin{align}
\frac{d\p}{d\tau}=\frac{dr}{d\tau}\frac{d\p}{dr}=\frac{\tau}{r}\frac{d\p}{dr}\ .
\end{align}
Denoting the point where the false vacuum occurs as $\p_{\text{fv}}$, we write down the initial conditions needed to find solutions to the equations of motion~\cite{Coleman:1977py}
\begin{align}
V(\p(r\rightarrow \infty))=V(\p_{\text{fv}})=0, \ \ \ \ \ \frac{d\p}{dr}\rvert_{r=0}=0\ .
\end{align}
The equations of motion for $\p$ and $\chi$ are~\cite{Rubakov:2002fi}
\begin{align}
   \label{eq:EoMrho0}
&\p''+\frac{D-1}{r}\p'-\p(\chi')^2-\frac{\pd V}{\pd\p}=0\ ,\\
   \label{eq:EoMchi0}
&\pd_r(r^{D-1}\p^2\chi')=0\ .
\end{align}
It is easy to solve \eqref{eq:EoMchi0} for $\chi'$,
\begin{align}
\chi'=\frac{L}{r^{D-1}\p^2}\ ,
\end{align}
where $L$ is a constant of motion, reminiscent of conserved angular momentum. Plugging $\chi'$ into the first equation of motion~\eqref{eq:EoMrho0}, we get
\begin{align}
&\p''+\frac{D-1}{r}\p'-\frac{L^2}{\p^{3}r^{2(D-1)}}-\frac{\pd V}{\pd\p}=0\ .
\end{align}
The Euclidean action then becomes
\begin{align}
S_E\ =\ &\frac{2 \pi^{\frac{D}{2}}}{\Gamma(1+\frac{D}{2})} \int_0^\infty dr \  r^{D-1}\bigg[\frac{1}{2}(\p')^2\: +\: \frac{1}{2}L^2\rho^{-2}r^{-2(D-1)}\:+\: V(\rho)\bigg]\ .
\end{align}
Interpreting the parameter $r$ as ``time'', we can draw the classical analogy of a particle sliding through a time-dependent effective potential given by
\begin{align}
V_{\text{eff}}(\p,r)\ =\ \frac{1}{2}L^2\p^{-2}r^{-2(D-1)}\: +\: \Big(\!-V(\p)\Big) \ ,
\end{align}
subject to a time-dependent drag force $\propto \p'/r$. This is analogous to Keplerian dynamics defined by a potential given by $-V(\p)$ with a time-dependent friction term. We would like to find the position from which the particle starts from rest, slides down the potential and slows down gradually to settle at the false vacuum at $r\rightarrow \infty$.

Setting $L=0$ is equivalent to eliminating the Goldstone bosons and only $\p(r)$ becomes relevant. The parameter $L = \chi' \p^{2}r^{D-1}$ must be constant at all ``times'' $r$, which implies that if $L$ is non-zero, then (for $D>1$) at $r=0$, at least one of $\chi'$ and $\p$ must be singular. The divergence at the origin is problematic, since it requires infinite energy for the field to tunnel through. As a classical analogue, we may imagine the situation of shrinking a rotating sphere to a point while conserving its angular momentum. This is impossible because that would require infinite rotational energy.

Moreover, there is an infinite barrier at $\rho=0$ due to the term in the effective potential that is proportional to $\p^{-2}$. Therefore, fields that move under the influence of potentials where the false vacuum occurs at $\p=0$ can never reach this value asymptotically without undergoing infinitely many oscillations. However, this gives an infinite value for the action. 

To avoid these problems, we can consider a minimal extension to the action that includes gravitational effects in curved spacetime, and introduce a hole at the origin of coordinates resembling a punctured spacetime.  For reasons mentioned above, we must also consider potentials where the false vacuum occurs at some positive value $\p>0$.

\subsection{Punctured curved spacetime}

As we include the Einstein-Hilbert action and allow for a general metric, the action in the presence of gravity (we choose to work in $D=4$) is given by~\cite{Lee:2012ug}
\begin{align}
S=&\int d^4x \sqrt{-g}\bigg[\frac{1}{2}g^{\mu\nu}\partial_\mu\rho\partial_\nu\rho+\frac{1}{2}g^{\mu\nu}\rho^2\partial_\mu\chi\partial_\nu\chi-V(\p)+\frac{R}{2\kappa}\bigg] \ ,
\end{align}
with $\kappa=8\pi G=8\pi /M_p^2$, where $G$ is Newton's gravitational constant and $M_p=1.22\times 10^{19}$ GeV  is the Planck mass. Upon Wick rotation of the time coordinate, we obtain the Euclidean action
\begin{align}
S_E=&\int d\tau d^3x \sqrt{g_E}\bigg[\frac{1}{2}g_E^{\mu\nu}\partial_\mu\rho\partial_\nu\rho+\frac{1}{2}g_E^{\mu\nu}\rho^2\partial_\mu\chi\partial_\nu\chi+V(\p)-\frac{R}{2\kappa}\bigg] \ .
\end{align}
We choose to work in an $O(4)$ symmetric configuration by constructing a general rotationally invariant metric defined by the line element~\cite{Coleman:1980aw,Branchina:2018xdh}
\begin{align}
ds^2=dr^2+a^2(r)d\Omega^2 \ .
\end{align}
The variation with respect to the metric gives rise to the Einstein equations
\begin{align}
    \label{eq:Grr}
G_{rr}&=3\bigg(\frac{(\partial_r a)^2-1}{a^2}\bigg)=\kappa\bigg(\frac{1}{2}\partial_r\rho\partial_r\rho+\frac{1}{2}\rho^2\partial_r\chi\partial_r\chi-V(\p)\bigg)\ ,\\
   \label{eq:Ricci}
\frac{R}{2\kappa}&=\frac{1}{2}g_E^{\mu\nu}\partial_\mu\rho\partial_\nu\rho+\frac{1}{2}g_E^{\mu\nu}\rho^2\partial_\mu\chi\partial_\nu\chi+2V(\p)\ ,
\end{align}
while the variations with respect to $\rho$ and $\chi$ yield
\begin{align}
&\rho''+3\frac{a'}{a}\rho'-\rho\chi'^2-\frac{dV}{d\p}=0 \ ,\\
&\chi'=L\rho^{-2}a^{-3} \ .
\end{align}
Combining these results, we get the following coupled differential equations:
\begin{align}
	\label{eq:EoMrho1}
&\rho''+3\frac{a'}{a}\rho'-\frac{L^2 }{\rho^{3}a^{6}}-\frac{dV}{d\p}=0\; ,\\
&a'^2=1+\frac{\kappa}{3}a^2\bigg(\hf \rho'^2+\hf L^2 \rho^{-2}a^{-6}-V(\p)\bigg)\; .
\end{align}
Again, we make the classical analogy of a particle moving under the influence of an effective potential of the form
\begin{align}
\label{eq:Veff2}
V_{\text{eff}}(\p(r),a(r))\ =\ \frac{1}{2}L^2\p^{-2}a^{-6}\: +\: \Big(\!-V(\p)\Big)\; ,
\end{align}
subject to a time-dependent drag force $\propto a'\p'/a$. The Euclidean action $S_E$ can now be written as
\begin{align}
S_E=2\pi^2&\int dr \bigg[a^3\bigg(\hf\p'^2+\hf L^2\p^{-2}a^{-6}+V(\p)\bigg)+\frac{3}{\kappa}\bigg(a^2a''+aa'^2-a\bigg)\bigg] \; .
\end{align}
Using~\eqref{eq:Ricci}, we may simplify the expression for the Euclidean action as
\begin{align}
S_E=-2\pi^2&\int dr\  a^3V(\p) \ .
\end{align}

Unlike the case in which gravity is absent, provided that $a(r=0)\neq 0$, nothing can prevent us from obtaining finite energy solutions that satisfy the boundary condition $\rho'(0)=0$. Moreover, as the field $\rho$ vanishes, the metric should become flat, which reads $a(r)-r\rvert_{r\rightarrow \infty}=\text{constant}$. We choose the following boundary conditions:
\begin{align}
&\p(r\rightarrow\infty)=\pfv \ , \ \ \ \ \ \p'(0)=0 \ ,\nonumber \\
&a(0)\equiv a_0\neq0\ ,\ \ \ \ \ a'(0)=0 \ .
\end{align}
Setting $L\neq 0$ is equivalent to giving the complex field ``angular momentum''. Since Goldstone fields do not appear in the potential, they will not contribute to the vacuum expectation value. However, their kinetic energy allows us to tunnel through the barrier to a field value $\p(0)\equiv \p_0$ where $V(\p_0)>V(\p_{\text{fv}})$.

The boundary condition $a'(0)=0$ ensures that there is no cusp in the metric, which results from the fact that the function $a^2(r)$ must be even when $r\rightarrow -r$ because of spherical symmetry. Consequently, we get an expression for $L$ in terms of $a_0$ and $\p_0$ that satisfies the condition
\begin{align}
L^2= 2\kappa^{-1}a_0^4\rho_0^2\Big(a_0^2V(\rho_0)-3\kappa^{-1}\Big)\;  .
\end{align}
But, since $L^2$ is non-negative, we must then necessarily have 
\begin{align}
a_0^2\: \geq\: 3/\kappa V(\rho_0)\: \geq\: 3/\kappa V_{\text{max}} \; ,
\end{align}
where $V_{\text{max}}$ is the height of the barrier. Consistent with this constraint, we parametrise 
$a_0$ as
\begin{align}
a_0^2\ =\ \frac{3}{\kappa V(\p_0)(1-\delta)}\ ,
\end{align}
where the parameter $\delta$ takes on values in the interval: $0\leq \delta <1$. This allows us to simplify the expression for $L$ as
\begin{align}
   \label{eq:L2}
L^2\ =\ 2\delta  a_0^6\p_0^2  \Vo \;  .
\end{align}

The parameter $a_0$ resembles the radius of the throat of a wormhole in spacetime. Choosing a value for this parameter of order $O(M_p)$ corresponds to a wormhole that is a few kilometers wide, comparable to sizes of typical blackholes in nature.

Since the function $a(r)$ is ever increasing, we can safely neglect drag forces as they are heavily suppressed by the size of the hole provided that we choose a large value for $a_0$, which is ensured if $O(\kappa V_{\text{max}})\ll 1$. As a consequence, $\frac{1}{2}L^2\p^{-2}a^{-6}+V(\p)$ is non-negative for all $r$, since otherwise the particle will overshoot the false vacuum. This implies by itself that the Hamiltonian of the theory, $T_{rr}=\hf \p'^2 + \hf L^2 \p^{-2}a^{-6}+V(\p)$, is positive definite, satisfying the null energy condition. Moreover, we expect that the vacuum would be very short-lived, as the value of the Euclidean action $S_E$ will be largely negative.

We use ~\eqref{eq:Veff2} to give a bound on the minimum energy required for the tunneling process. The condition $V_{\text{eff}}(\p,r=0)> -V(\pfv)$ must be satisfied, which translates to $(1-\delta)V(\p_0)<V(\pfv)$. This sets a lower bound on $a_0$ for which solutions that satisfy the boundary conditions exist,

\begin{align}
	\label{eq:amin}
a_0 \ > a_0^{\text{min}} \equiv \sqrt{\frac{3}{\kappa V(\pfv)}}\ .
\end{align}

Hence, $V(\pfv)$ must be positive. 

\section{Numerical analysis}\label{sec:Numerics}

In this section, we will numerically find finite-action solutions satisfying the appropriate boundary conditions for different potentials. In particular, we consider the Coleman model examining both ways in which the false and true vacua are arranged in the potential; (i) $\pfv<\ptv$, (ii) $\pfv>\ptv$. Finally, we analyse a Fubini potential with a non-vanishing mass term.

After fixing the parameter $a_0$ to a specific value, we use the bisection method to determine the unique tunnelling point $\p_0$ from which the particle evolves and asymptotes the false vacuum. Starting from any other value for $\p_0$, this will either overshoot or undershoot the false vacuum. But, after undergoing many iterations, we will be able to determine the value which satisfies the boundary conditions to a great accuracy.

\subsection{Coleman potential with $\pfv<\ptv$}

\begin{figure}[ht!]
\begin{center}
\includegraphics[width=0.5\textwidth]{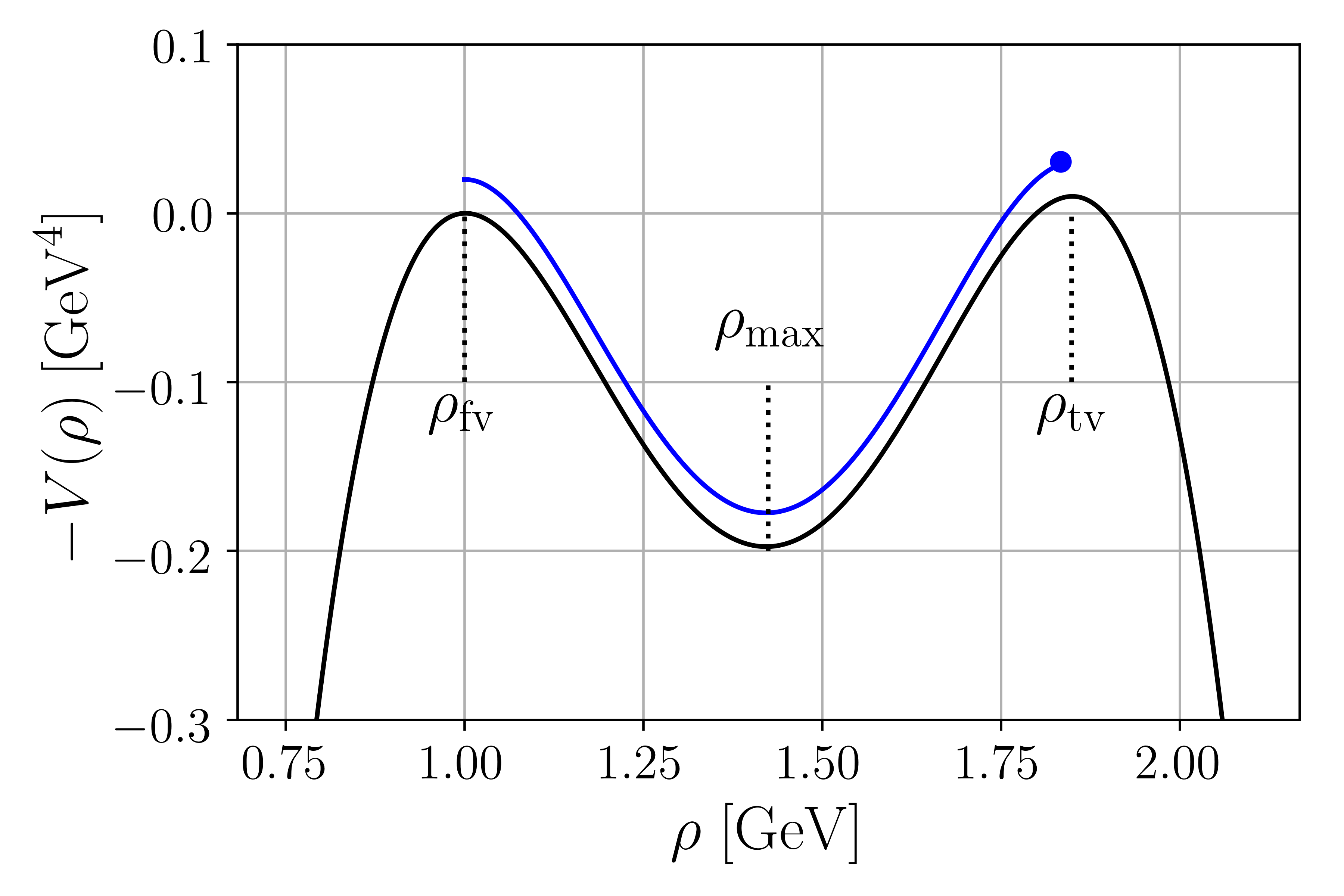}
\caption{The figure displays a Coleman potential ($\pfv<\ptv$) with $\pfv=\psh-\sqrt{\mu^2/\lambda}$ and $\ptv=\psh+\sqrt{\mu^2/\lambda}$  determined by the parameters $\mu=3 \ \text{GeV}, \lambda=50, \epsilon=0.01 \ \text{GeV}^4$ and $\psh=\sqrt{\mu^2/\lambda}+\Delta\p $, with $\Delta \p=1 \  \text{GeV}$. The blue line shows the trajectory of the bounce solution , in the case of $L\neq 0$, for a fixed $V_0=0.01, a_0=4.26\times 10^{19} \ \text{GeV}^{-1}$ with the blue circle indicating the tunnelling point $\p_0$. }
\label{fig:ColemanA}
\end{center}
\end{figure}

Let us consider a shifted Coleman potential, which is displayed in the left panel of Fig.~\ref{fig:ColemanA} and is given by~\cite{Coleman:1977py}
\begin{align}
   \label{eq:VColemanA}
&V(\p)\ =\ V_+(\p)\: -\: \frac{\epsilon}{2\sqrt{\mu^2/\lambda}}\,\Big[(\p-\psh)+\sqrt{\mu^2/\lambda}\,\Big]+V_0\; ,
\end{align} 
where 
\begin{align}
   \label{eq:Vplusrho}
&V_{+}(\p)\ =\ \frac{\lambda}{8}\bigg((\p-\psh)^2-\frac{\mu^2}{\lambda}\bigg)^2\;,
\end{align}
with
\begin{align}
&\pfv\equiv\psh-\sqrt{\frac{\mu^2}{\lambda}}\;,\qquad \ptv\equiv\psh+\sqrt{\frac{\mu^2}{\lambda}}\; .
\end{align}
In the above, the constants $\mu,\lambda, V_0$ and $\epsilon$ are all positive, with the latter being very small compared to the height of the potential $V(\p_{\text{max}})$. We choose $\psh>\sqrt{\mu^2/\lambda}$, such that all minima lie on positive values of the field $\p$. We can easily see that $ V(\p_{\text{fv}})=V_0$ and $V(\ptv)=-\epsilon+V_0$. To parametrise the potential, we choose the values $\mu=3 \ \text{GeV}, \lambda=50, \epsilon=0.01 \ \text{GeV}^4$, and $\psh=\sqrt{\mu^2/\lambda}+\Delta\p $, with $\Delta \p=1 \  \text{GeV}$. 

For a potential $V(\p)$ given in~\eqref{eq:VColemanA}, we obtain a solution in the so-called thin wall approximation, when $L=0$. The field $\rho$ starts rolling at a value which is very close to the true vacuum $\ptv$ where it settles until a very large time $r=R$, when the time-dependent friction term becomes negligible. It then starts sliding rapidly down the potential and asymptotes the false vacuum $\pfv$. The profile of $\rho$ in the thin wall approximation may conveniently be expressed as
\begin{equation}
\p(r)\ =\ 
\begin{cases} 
      \ptv & r\ll R\\
       \sqrt{\frac{\mu^2}{\lambda}}\tanh\bigg(\hf\mu(r-R)\bigg)+\psh& r\simeq R\\
      \pfv & r\gg R
   \end{cases}
\end{equation}

The calculations for the Euclidean action corresponding to the thin wall solution were laid out in detail by Coleman and De Luccia~\cite{Coleman:1980aw}. We quote their results below 
\begin{align}
S_E=\frac{12\pi^2}{\kappa^2}\bigg[&V^{-1}(\ptv)\bigg(1-\frac{\kappa}{3}a^2(R)V(\ptv)\bigg)^{3/2} -V^{-1}(\pfv)\bigg(1-\frac{\kappa}{3}a^2(R)V(\pfv)\bigg)^{3/2}\nonumber\\& -a(R)\rightarrow a_0\bigg]+2\pi^2a^3(R)S_+  \ ,
\end{align}
where $S_+=\int_{\ptv}^{\pfv}  d\p \ [2V_+(\p)]^{-1/2}=\mu^3/3\lambda $. The validity of the approximation relies on satisfying the condition $\mu^4/\epsilon\lambda \gg 1$. Minimising the action with respect to $a(R)$ gives~\cite{Coleman:1980aw}
\begin{align}
a(R)=\frac{3S_+}{\epsilon[1-(3S_+/2\epsilon\Lambda)^2]} \ ,
\end{align} 
where $\Lambda=(\kappa\epsilon/3)^{-1/2}$. Hence, in the limit $a_0\rightarrow 0$, the Euclidean action takes the simple form~\cite{Coleman:1980aw}
\begin{align}
S_E=\frac{27\pi^2S_+^4}{2\epsilon^3[1-(3S_+/2\epsilon\Lambda)^2]^2} \ .
\end{align} 
 
We start our analysis by choosing values for the parameter $a_0$, and then numerically find finite-action solutions that asymptote the false vacuum corresponding to a unique tunnelling point $\p_0$, which in turn determines the value of $L^2$ and satisfies the boundary conditions.

\begin{figure}[ht!]
\begin{center}
 \begin{minipage}[b]{0.49\textwidth}
    \includegraphics[width=\textwidth]{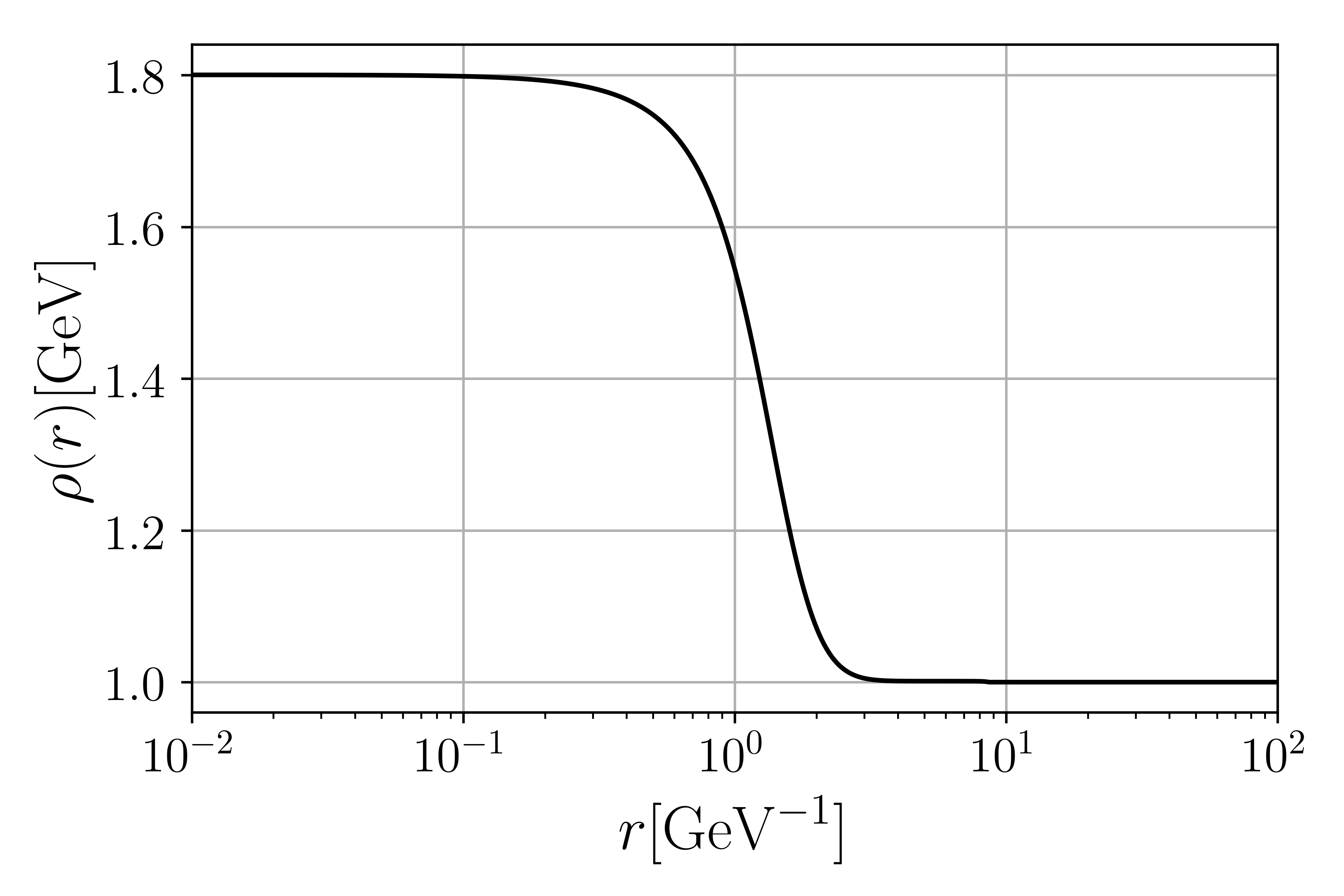}   
  \end{minipage}
  \hfill
  \begin{minipage}[b]{0.49\textwidth}
    \includegraphics[width=\textwidth]{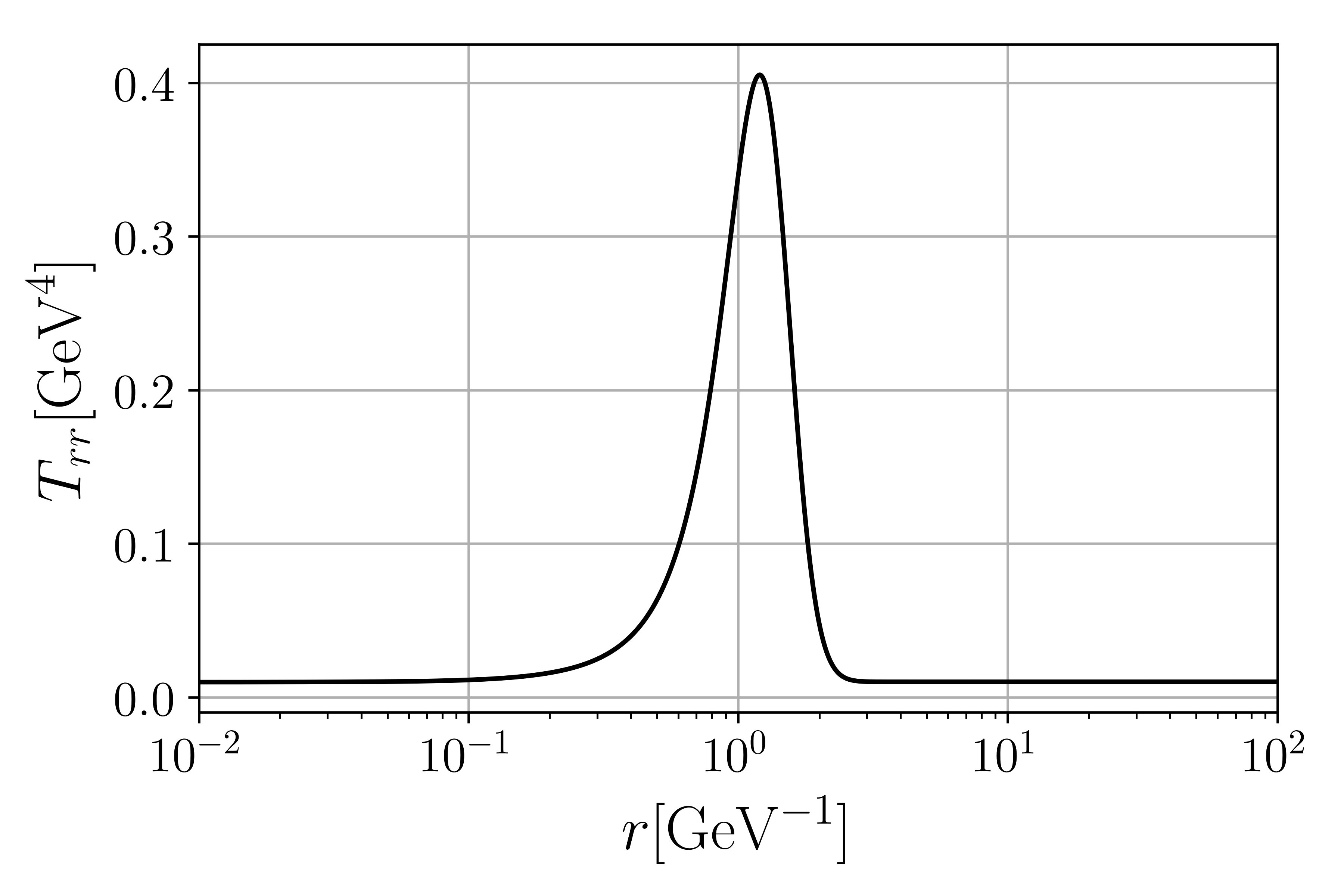}
  \end{minipage}
\caption{The bounce solution $\p(r)$ for $V_0=0.01 \ \text{GeV}^4, a_0=4.26\times10^{19} \  \text{GeV}^{-1}, \p_0=1.8002 \ \text{GeV}$, under a Coleman potential $(\pfv<\ptv)$ parametrised by $\mu=3 \ \text{GeV}, \lambda=50, \epsilon=0.01 \ \text{GeV}^4$ and $\psh=\sqrt{\mu^2/\lambda}+\Delta\p $, with $\Delta \p=1 \  \text{GeV}$. }
\label{fig:rhoColemanA}
\end{center}
\end{figure}

The term $L^2\p^{-3}a^{-6}$ in~\eqref{eq:EoMrho1} accelerates the field $\p$ towards positive values, which means that it drives the $\rho$ field away from the false vacuum. Therefore, it will always undershoot the false vacuum unless it starts sliding near the true vacuum. The field climbs up the potential to reach the true vacuum before sliding down the potential and asymptoting the false vacuum. This is visualised by the nucleation of a bubble where the interior resides in the true vacuum and outside the bubble the Universe will occupy the false vacuum state. In the right panel of~Fig.~\ref{fig:rhoColemanA}, we see that the relevant Hamiltonian $T_{rr}$ is positive definite, satisfying the null energy condition.

\setlength{\tabcolsep}{12pt}
\begin{table}[]
\centering

\begin{tabular}{ccccc}
  \hline
$V_0[\text{GeV}^{4}]$&$a^{\text{min}}_0[\text{GeV}^{-1}]$&$\rho_0[\text{GeV}]$ & $L^2$ & $B$ \\
\hline
0.01&4.26$\times 10^{19}$ & 1.8002 & 3.08 $\times10^{114} $& $-2.64\times10^{59}$  \\ 
0.1&1.35$\times 10^{19}$ & 1.8037 & 2.43 $\times10^{112} $& $-8.31\times10^{57}$  \\ 
0.25&8.52$\times 10^{18}$ & 1.8098 & 3.82$\times10^{111} $& $-2.08\times10^{57}$  \\ 
0.5&6.02$\times 10^{18}$ & 1.8101 & 9.49$\times10^{110} $& $-7.20\times10^{56}$  \\ 
1&4.26$\times 10^{18}$ & 1.8167 & 5.82$\times10^{110} $& $-2.54\times10^{56}$  \\
  \hline
\end{tabular}
\caption{Numerical estimates of the Euclidean action for different values of $V_0, a_0,\p_0$ and $L$, for a Coleman potential ($\pfv<\ptv$) determined by the parameters $\mu=3\,\text{GeV}, \lambda=50, \epsilon=0.01\,\text{GeV}^4$ and $\psh=\sqrt{\mu^2/\lambda}+\Delta\p$, with $\Delta \p=1\,\text{GeV}$.  For comparison, the Euclidean action corresponding to the $L=0$ case is $B=+1.40\times 10^6$.}
\label{tab:ColemanA}
\end{table}

In Table~\ref{tab:ColemanA}, we give numerical estimates of the Euclidean action for different values of $a_0,\p_0$ and $L$, for the Coleman potential with $\pfv<\ptv$. We display the value of the critical hole size $a^{\text{min}}_0$ and the corresponding $\p_0,L,B$, for which solutions exist. We obtain very large and negative values for~$B$ ranging from $-10^{56}$ to $-10^{59}$, which implies that the presence of Goldstone modes can make the vacuum highly unstable. For comparison, we note that the Euclidean action corresponding to the $L=0$ case is $B=+1.40\times 10^6$, predicting a vacuum which is very stable.

\subsection{Coleman potential with $\pfv>\ptv$}


\begin{figure}[h!]
\begin{center}
 \includegraphics[width=0.5\textwidth]{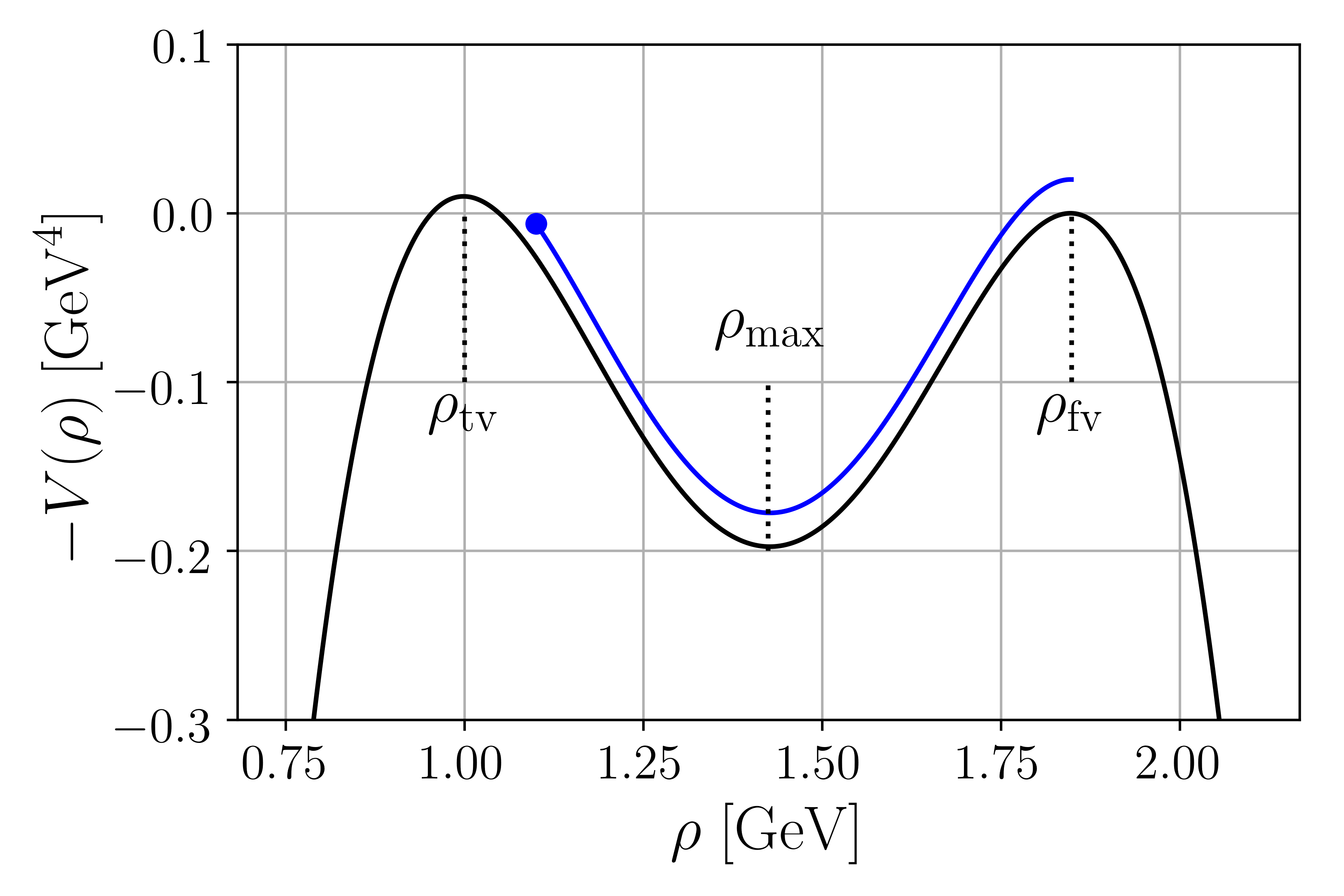}
\caption{The figure displays a Coleman potential ($\pfv>\ptv$) with $\pfv=\psh+\sqrt{\mu^2/\lambda}$ and $\ptv=\psh-\sqrt{\mu^2/\lambda}$  determined by the parameters $\mu=3 \ \text{GeV}, \lambda=50, \epsilon=0.01 \ \text{GeV}^4$ and $\psh=\sqrt{\mu^2/\lambda}+\Delta\p $, with $\Delta \p=1 \  \text{GeV}$. The blue line shows the trajectory of the bounce solution , in the case of $L\neq 0$, for a fixed $a_0=4.26\times 10^{19}\  \text{GeV}^{-1}$ with blue circle indicating the tunnelling point $\p_0$.}
\label{fig:ColemanB}
\end{center}
\end{figure}
As shown in the left panel of Fig.~\ref{fig:ColemanB}, we now consider a potential, for which $\pfv>\ptv$, given by
\begin{align}
V(\p)&\ =\ V_+(\p)\: +\: \frac{\epsilon}{2\sqrt{\mu^2/\lambda}}\,\Big[(\p-\psh)-\sqrt{\mu^2/\lambda}\,\Big]+ V_0\; , 
\end{align}
such that
\begin{align}
\pfv&\ \equiv\ \psh+\sqrt{\frac{\mu^2}{\lambda}}\; ,\qquad   \ptv\ \equiv\ \psh-\sqrt{\frac{\mu^2}{\lambda}}\; ,
\end{align}
where $V_+(\p)$ is defined in~\eqref{eq:Vplusrho}.  The term $L^2\p^{-3}a^{-6}$ in~\eqref{eq:EoMrho1} now accelerates the field towards the false vacuum $\pfv$. This means that the particle slides down the potential immediately after tunnelling to a point just below the true vacuum, where the potential at that point is positive, and rapidly asymptotes the false vacuum. 

\begin{figure}[h!]
\begin{center}
 \begin{minipage}[b]{0.49\textwidth}
    \includegraphics[width=\textwidth]{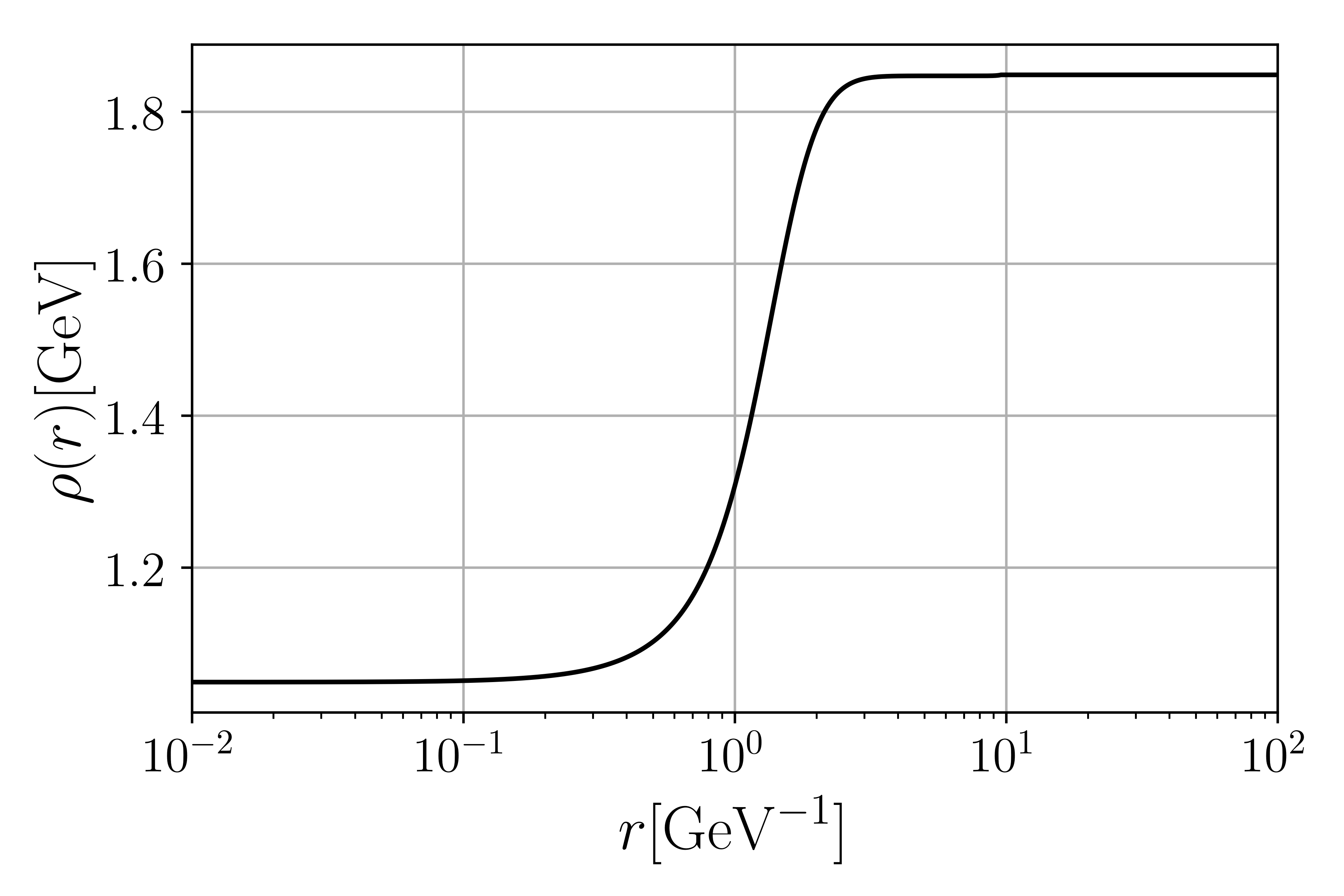}   
  \end{minipage}
  \hfill
  \begin{minipage}[b]{0.49\textwidth}
    \includegraphics[width=\textwidth]{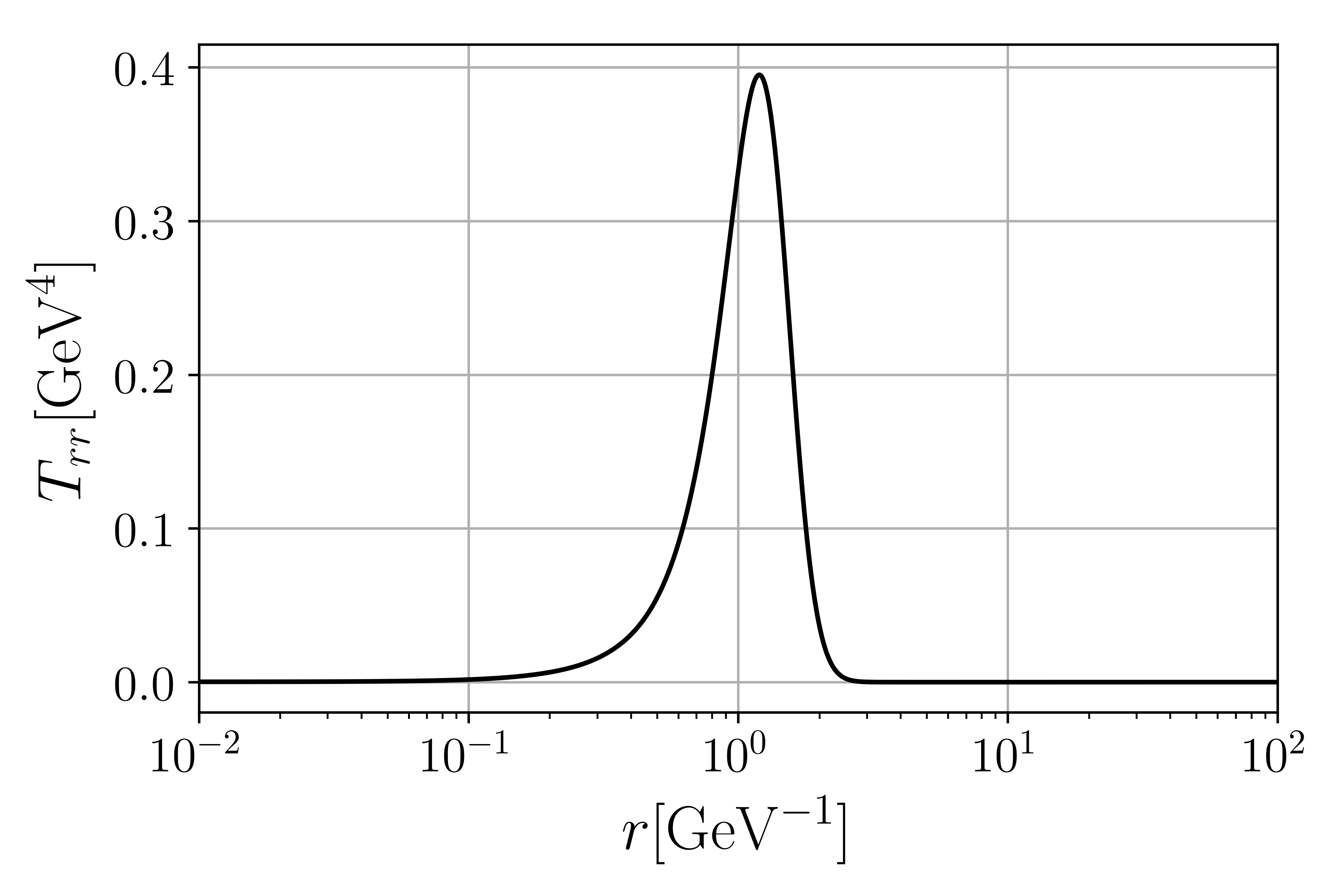}
  \end{minipage}
\caption{The bounce solution $\p(r)$ for $V_0=0.01 \ \text{GeV}^4, a_0=4.26\times10^{19} \  \text{GeV}^{-1}, \p_0=1.0496 \ \text{GeV}$, under a Coleman potential $(\pfv>\ptv)$ determined by the parameters $\mu=3 \ \text{GeV}, \lambda=50, \epsilon=0.01 \ \text{GeV}^4$ and $\psh=\sqrt{\mu^2/\lambda}+\Delta\p $, with $\Delta \p=1 \  \text{GeV}$.}
\label{fig: rhoColemanB}
\end{center}
\end{figure}

As exhibited in Table~\ref{tab:ColemanB}, the lifetime of the vacuum becomes now longer than the one in the Coleman case with $\pfv<\ptv$. But, the vacuum is equally very unstable, since $B$ ranges from $-10^{56}$ to $-10^{59}$. Comparing these results to the ones obtained earlier highlights that the order of which the false and true vacua are placed within the potential makes a significant difference, as tunnelling rates differ considerably. This asymmetry is a consequence of the existence of Goldstone modes which drive the $\rho$ field in only one direction. For the usual scenario with $L=0$, the profiles of solutions and the tunnelling rates for the two potentials are similar. Instead, when $L\neq 0$, the relative field value between the vacuum states does matter.

\setlength{\tabcolsep}{12pt}
\begin{table}[]
\centering
\begin{tabular}{ccccc}
\hline
$V_0[\text{GeV}^{4}]$&$a^{\text{min}}_0[\text{GeV}^{-1}]$&$\rho_0[\text{GeV}]$ & $L^2$ & $B$ \\
\hline
0.01&4.26$\times 10^{19}$ & 1.0496 & 8.00 $\times10^{114} $& $-2.65\times10^{59}$  \\ 
0.1&1.35$\times 10^{19}$ & 1.0589 & 8.29 $\times10^{112} $& $-8.45\times10^{57}$  \\ 
0.25&8.52$\times 10^{18}$ & 1.0710 & 1.30$\times10^{112} $& $-2.17\times10^{57}$  \\ 
0.5&6.02$\times 10^{18}$ & 1.0875 & 3.23$\times10^{111} $& $-7.79\times10^{56}$  \\ 
1&4.26$\times 10^{18}$ & 1.1132 & 8.05$\times10^{110} $& $-2.83\times10^{56}$  \\ 
\hline
\end{tabular}
\caption{The same as in Table~\ref{tab:ColemanA}, using the same input parameters,
but for a Coleman potential with $\pfv>\ptv$.}
\label{tab:ColemanB}
\end{table}

\subsection{Fubini potential with a mass term}

\begin{figure}[h!]
\begin{center}
 \includegraphics[width=0.5\textwidth]{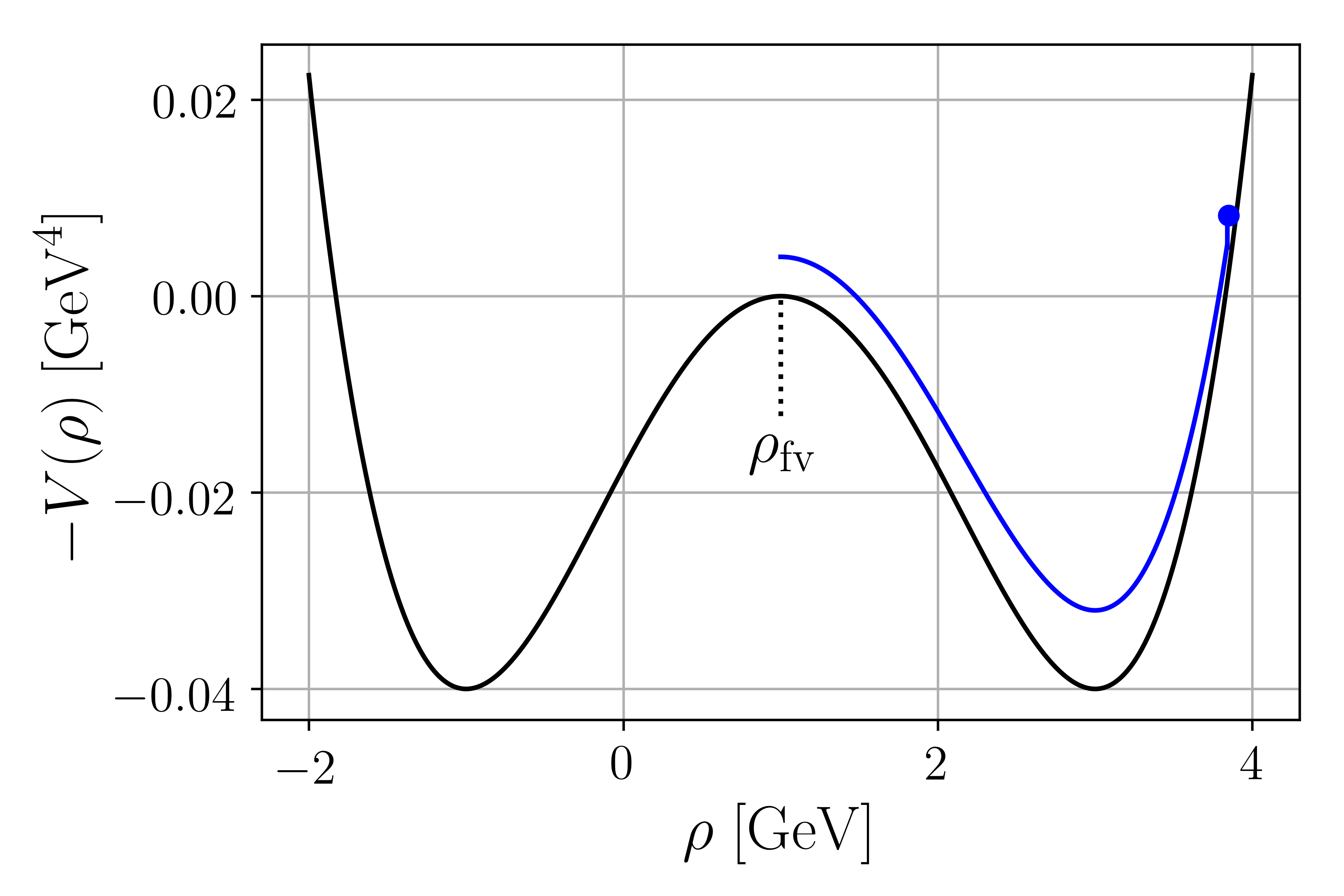}
\caption{The figure displays a massive Fubini potential parametrised by $\lambda=-0.01$, $m=0.2$ GeV, and $\pfv=1$ GeV. The blue line shows the trajectory of the bounce solution , in the case of $L\neq 0$, for a fixed $a_0=4.26\times 10^{19}\  \text{GeV}^{-1}$ with the blue circle indicating the tunnelling point $\p_0$. }
\label{fig:Fubini}
\end{center}
\end{figure}

\begin{figure}[h!]
\begin{center}
 \begin{minipage}[b]{0.49\textwidth}
    \includegraphics[width=\textwidth]{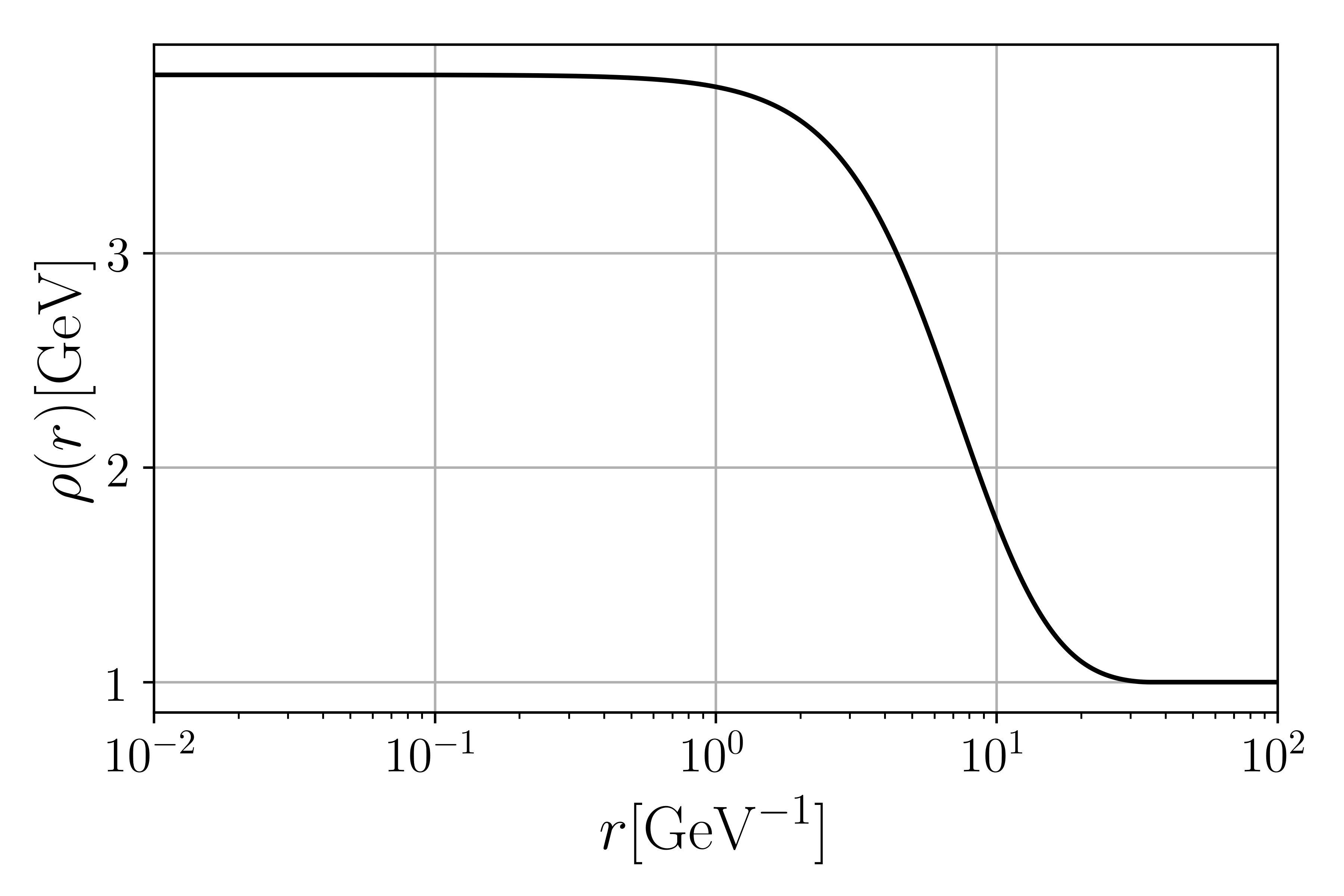}   
  \end{minipage}
  \hfill
  \begin{minipage}[b]{0.49\textwidth}
    \includegraphics[width=\textwidth]{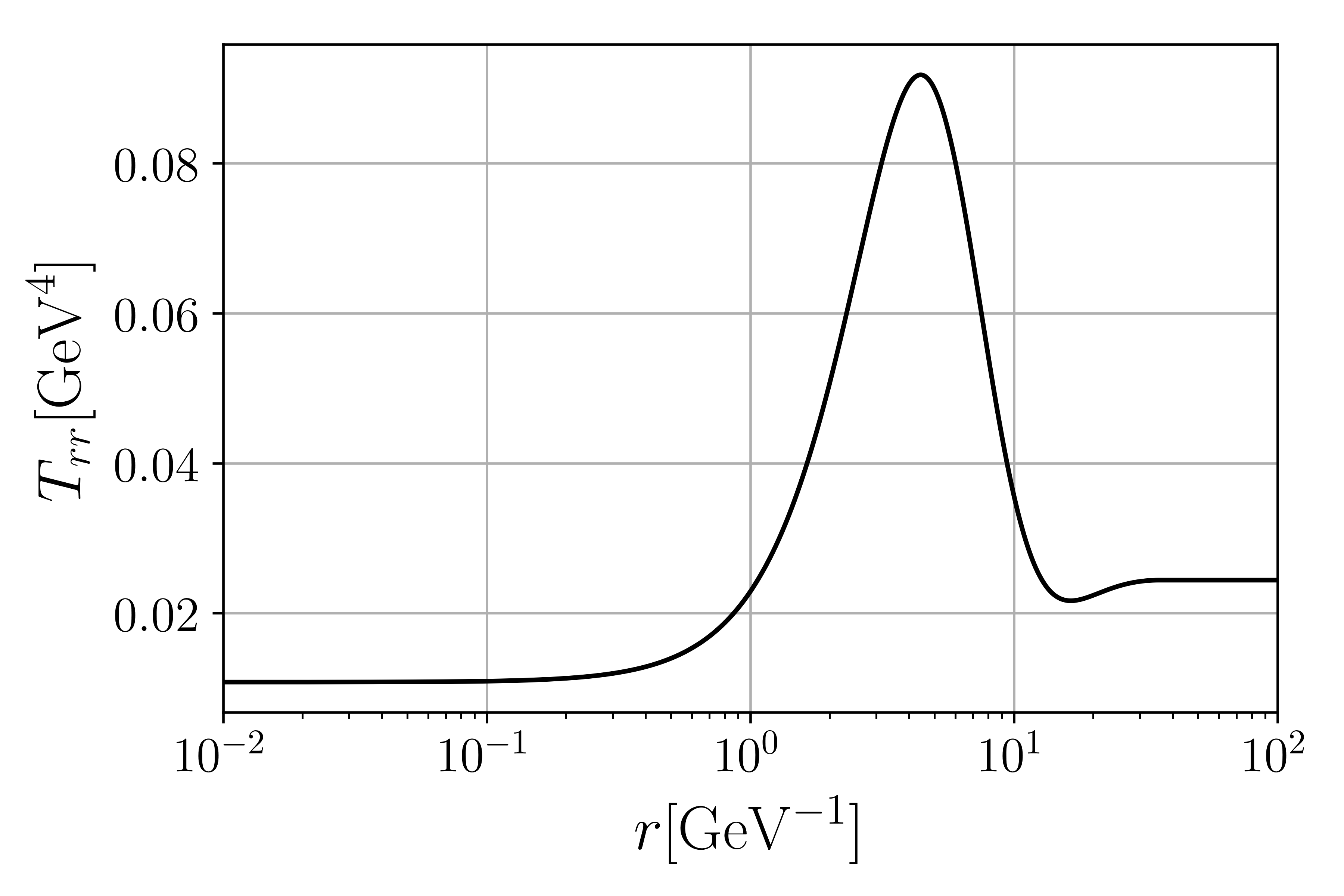}
  \end{minipage}
\caption{The bounce solution $\p(r)$ for $V_0=0.01 \ \text{GeV}^4, a_0=4.26\times10^{19} \  \text{GeV}^{-1}, \p_0=3.8300 \ \text{GeV}$,  under a Fubini potential determined by the parameters $\lambda=-0.01$, $m=0.2$ GeV, and $\pfv=1$ GeV.}
\label{fig:rhoFubini}
\end{center}
\end{figure}

In this subsection, as shown  in the left panel of Fig.~\ref{fig:Fubini}, we will analyze a shifted Fubini potential with a mass term \cite{deAlfaro:1976vlx}, which has the form
\begin{align}
V(\p)=\frac{\lambda}{4}(\p-\pfv)^4+\hf m^2(\p-\pfv)^2 + V_0\ ,
\end{align}
with $\lambda<0$. In this case, the potential has no true vacuum and is unbounded from below. We choose to parametrise the potential by setting $\lambda=-0.01$, $m=0.2$ GeV, and $\pfv=1$ GeV. For $L=0$, using arguments of scale invariance~\cite{Affleck:1980mp}, it is possible to show that no solutions exist for a non-vanishing $m$ in flat spacetime. The friction term will always prevent the field from climbing up the hill and reaching the false vacuum at infinity. Therefore, one needs extra operators to find tunnelling solutions. For example, it can be achieved by adding to the potential a term $\propto \p^6/\Lambda^2$ suppressed by some new scale $\Lambda$~\cite{Affleck:1980mp}, or by including gravitational effects and introducing a cosmological constant \cite{Lee:2012ug,Lee:2014ula,Rubakov:1999ir}. 
In punctured curved spacetime with $L \neq 0$, the additional the rotational kinetic energy of the Goldstone field would provide the necessary energy to overcome drag forces. Thus, we are able to obtain a finite-action tunnelling solution for such a potential, which is given in  Fig.~\ref{fig:rhoFubini}. Our numerical estimates are presented in Table~\ref{tab:Fubini}, predicting a very unstable vacuum for certain values of $a_0$ and $L$. \\

\setlength{\tabcolsep}{15pt}
\begin{table}[!ht]
\centering
\begin{tabular}{ccccc}
\hline
$V_0[\text{GeV}^{4}]$&$a^{\text{min}}_0[\text{GeV}^{-1}]$&$\rho_0[\text{GeV}]$ & $L^2$ & $B$ \\
\hline
0.01&4.26$\times 10^{19}$ & 3.8300 & 3.44 $\times10^{116} $& $-4.05\times10^{59}$  \\ 
0.1&1.35$\times 10^{19}$ & 3.8444 & 3.47 $\times10^{114} $& $-1.23\times10^{58}$  \\ 
0.25&8.52$\times 10^{18}$ & 3.8677 & 5.61$\times10^{113} $& $-2.93\times10^{57}$  \\ 
0.5&6.02$\times 10^{18}$ & 3.9046 & 1.43$\times10^{113} $& $-9.50\times10^{56}$  \\ 
1&4.26$\times 10^{18}$ & 3.9725 & 3.70$\times10^{112} $& $-2.83\times10^{56}$  \\ 
\hline
\end{tabular}
\caption{Numerical estimates of the Euclidean action for different input values of $V_0, a_0,\p_0$ and $L$, for a Fubini potential parametrised by $\lambda=-0.01$, $m=0.2$ GeV, and $\pfv=1$ GeV. }
\label{tab:Fubini}
\end{table}

As shown in Fig.~\ref{fig:Fubini}, the solution given describes a field that climbs up the potential then starts sliding down from a turning point and asymptotes the false vacuum, exactly like the Coleman case $\pfv<\ptv$ as the force associated with Goldstone modes drive the field away from the false vacuum.

We have shown that it is possible to find tunnelling solutions for the massive Fubini potential in curved spacetime which were deemed unattainable in flat spacetime. We can extend this result to any potential, because we can always compute the kinetic energy of the Goldstone field that is needed to overcome the time-dependent friction force which dies out as $r\to \infty$.

Another possible implication of Goldstone modes is the {\em reverse} tunnelling from a true vacuum to a false vacuum within a potential, since the kinetic energy of these modes can make up for the negative energy difference between the two minima.

\section{Conclusions}\label{sec:Conclusions}

We have studied the effects of Goldstone modes on the decay of the vacuum in flat spacetime, for $L\neq 0$, and concluded the impossibility of finding any physically viable solution due to a divergence at the origin. However, in punctured spacetime, it is possible to find finite-action solutions depending on the size of the hole $a_0$, which uniquely determines the value of $L$ and the position to which the field arrives after tunnelling $\p_0$. In particular, we obtain very large and negative values for the Euclidean action (cf.~Tables~\ref{tab:ColemanA}--\ref{tab:Fubini}). 
As a consequence, the rates of vacuum decay get drastically enhanced by many orders of magnitude, once the conditions~\eqref{eq:L2} and \eqref{eq:amin} are met, rendering the vacuum for such topological configurations highly unstable.

Other implications of the existence of such Goldstone modes include the possibility of reverse tunnelling to a field value that corresponds to higher potential energy as the rotational kinetic energy of the Goldstone fields would make up for the negative energy difference, something that is not possible with real scalar fields. Moreover, we have found that the order in which the false and true vacua are placed within the potential is significant to the lifetime of the vacuum and the profile of the resultant bubbles. This asymmetry originates from the fact that Goldstone modes accelerate the field $\p$ only towards higher field values. We were also able to obtain finite-action solutions corresponding to potentials which are deemed unsolvable in flat spacetime.

For such solutions to occur in nature, we must assume the existence of primordial wormholes, which are yet to be observed. Our results can be further improved by calculating contributions from Goldstone modes to the value of the prefactor $A$ in the decay rate. These decay mechanisms can explain different cosmological phenomena. For instance, one can describe inflationary scenarios using scalar fields which settled in short-lived false vacua \cite{Linde:2007fr}.

After tunnelling, the Universe will settle in a new vacuum changing the value of the VEV and leading to phase transitions. In the future, we might be able to detect signatures for such process in the form of gravitational waves. The theoretical framework for predicting the shape of the power spectra is a work in progress \cite{Caprini:2019egz}. The ESA is planning to build a laser interferometer and is scheduled to launch into space in the early 2030s under the LISA project,  enabling us to probe low frequency ranges which are typical of gravitational waves generated from these cosmological phase transitions.

\subsection*{Acknowledgments}
\vspace{-3mm}
We would like to thank Fedor Bezrukov for insightful comments. The work of AP is supported in part by the Lancaster-Manchester-Sheffield Consortium for Fundamental Physics, under STFC Research Grant No. ST/P000800/1. The work of MH is supported by UKSACB.


\end{document}